\documentclass[a4paper,11pt]{article}
\usepackage{pos}
\usepackage{graphicx}
\usepackage{graphbox}

\newcommand{\fbi}{fb$^{-1}$}
\def\nt{\tilde{\chi}^0}
\def\ch{\tilde{\chi}^\pm}

\usepackage{lineno}
\title{Digging deeper into SUSY parameter space with the CMS experiment}
\author*[a]{Sezen Sekmen, for the CMS Collaboration}

\affiliation[a]{Center for High Energy Physics, Kyungpook National University \\ Daegu, South Korea}

\emailAdd{ssekmen@cern.ch}

\abstract{The classic searches for supersymmetry have not given any strong indication for new physics.  Therefore CMS is designing dedicated searches to target the more difficult and specific supersymmetry scenarios. This contribution present three such recent searches based on 13 TeV proton-proton collisions recorded with the CMS detector in 2016, 2017 and 2018: a search for heavy gluinos cascading via heavy next-to-lightest neutralino in final states with boosted Z bosons and missing transverse momentum; a search for compressed supersymmetry in final states with soft taus; and a search for compressed, long-lived charginos in hadronic final states with disappearing tracks.}

\FullConference{%
  40th International Conference on High Energy physics - ICHEP2020\\
  July 28 - August 6, 2020\\
  Prague, Czech Republic (virtual meeting)
}

\begin{document}
\maketitle

% Parallel talk: 6 pages, including the title and reference pages.

%\section{Introduction}

The Compact Muon Solenoid (CMS) Experiment~\cite{Chatrchyan:2008aa} at the Large Hadron Collider (LHC) has collected an unprecedented 137 \fbi of data with proton-proton collisions at a center-of-mass energy of 13 TeV, which is continuously explored for traces of supersymmetry in a wide variety of searches.  As of 2020, classical searches, such as those looking for gluinos and squarks in inclusive SUSY final states with a large number of search bins, looking for top squarks in hadronic, single lepton or dilepton final states, or looking for charginos or neutralinos in single lepton, dilepton or trilepton final states have not yet observed a deviation from the standard model (SM), and excluded parts of the SUSY parameter space.  However, SUSY can still be realized in many alternative well-motivated ways in hidden, remote corners of the vast, multi-dimensional SUSY parameter space, to which the more standard and inclusive searches may not be sensitive.  Nowadays, CMS is enriching its physics program by an increasing diversity of dedicated searches to probe such corners and enhance the chances of discovery.

One example of a special scenario with a final state difficult to observe is that of a compressed mass spectrum where masses of two accessible SUSY partners are very close to each other.  Here, decays of the heavier particle lead to final states with low momentum (soft) objects and low missing transverse momentum.  Such final states are explored by searches that use soft objects and a high momentum initial state radiation jet.  Another consequence of compressed spectra can be long-lived particles, for which an increasing number of searches are being developed.  On the opposite end, there are scenarios with high mass SUSY partners and high mass differences,  for which several searches featuring objects with high Lorentz boost, leading to merged decay products, and thus substructure, are being designed.  Moreoever, there are dedicated searches for direct production of sleptons or staus, which are hard to access due to low cross sections and challenges in triggering.  Cascade decays with Higgs boson are also explored by explicitly reconstructing the Higgs boson and incorporating it into multi-object SUSY final states.  Additionally, signatures with special combinations of objects predicted by certain SUSY scenarios, such as $\gamma + b$ jets or $\gamma + $ lepton are investigated.  Besides all these searches, which are mainly targeting $R$-parity conserving models, a whole suite of analyses targeting variations of $R$-parity violating SUSY scenarios exist or are in progress.  This contribution presents 3 examples of recent non-classical SUSY searches based on CMS data collected in 2016, 2017 and 2018, aiming to dig deeper into the SUSY parameter space.

%%%%

%\section{Boosted $ZZ + p_T^{miss}$}

{\bf Boosted $ZZ + p_T^{miss}$ search: }
The first search is targeting a scenario motivated by naturalness, where pair-produced gluinos with $\sim$2-3 TeV mass decay cascading via a massive $\nt_2$ to a light $\nt_1$ and a $Z$ boson~\cite{Sirunyan:2020zzv}.  The large mass difference between $\nt_2$ and $\nt_1$ give the $Z$ bosons a large Lorentz boost.  The signature for a boosted $Z$ boson candidate is a wide-cone jet having a measured mass compatible with the $Z$ boson mass.  The analysis, performed on 137 fb$^{-1}$ of 13~TeV data, selects events with 0 leptons, $\ge 2$ $Z$ bosons, $\ge 2$ jets, missing transverse momentum $p_T^{miss} > 300$~GeV and hadronic transverse momentum $H_T > 400$~GeV.  The $Z$ boson candidates are selected among anti-$k_T$ jets with a size parameter of 0.8 (AK8 jets), and are required to have $p_T > 200$~GeV and a mass of $40$~GeV$ < m_{AK8jet} < 140$~GeV.  The 2nd highest $p_T$ $Z$ boson should be separated from any $b$-jet by an angular distance of $\Delta R(Z_2, b) > 0.8$ to eliminate backgrounds.

\begin{figure}[htbp]
\centering
\includegraphics[align=c, width=0.38\textwidth]{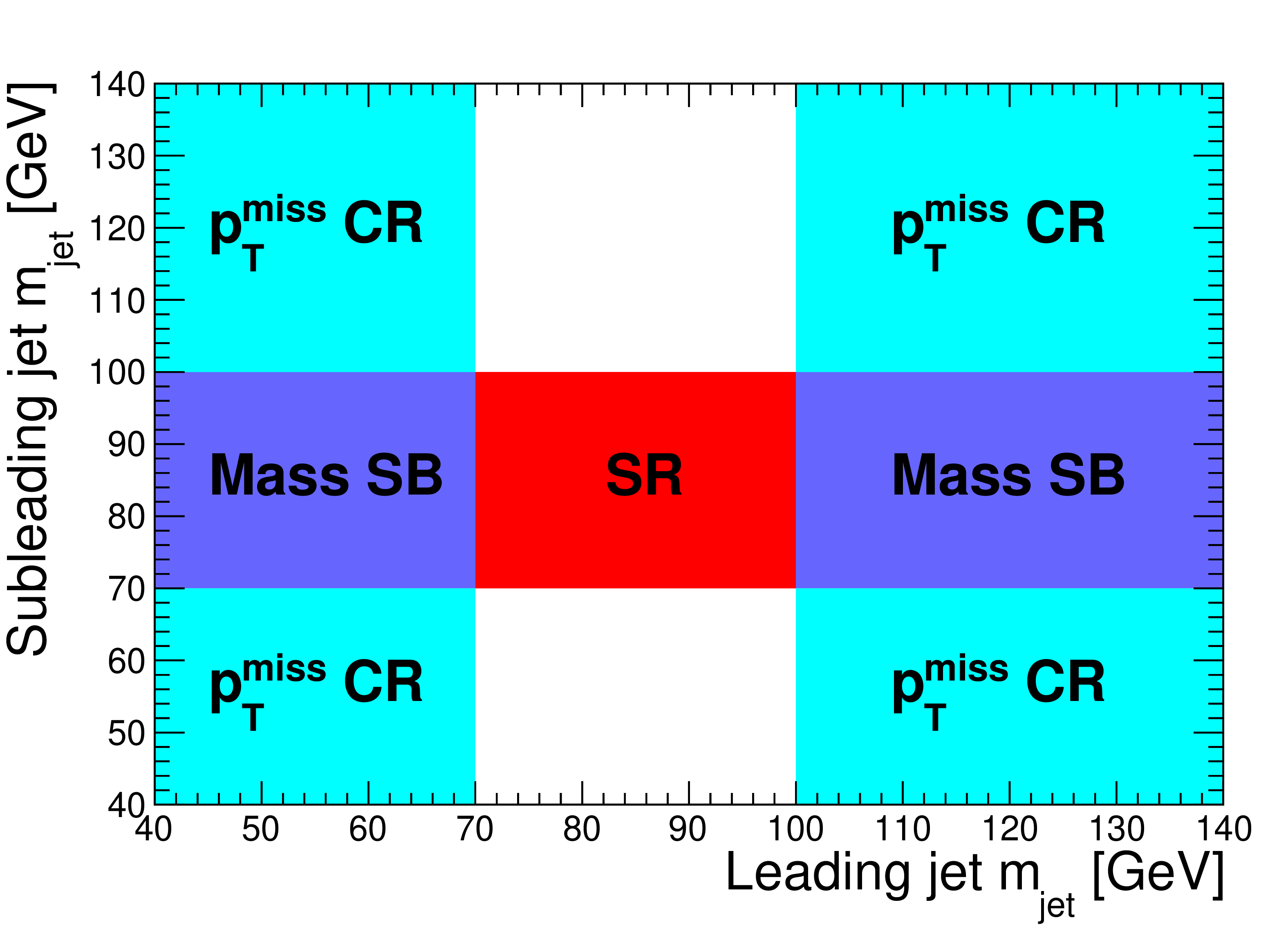} \qquad
\includegraphics[align=c, width=0.38\textwidth]{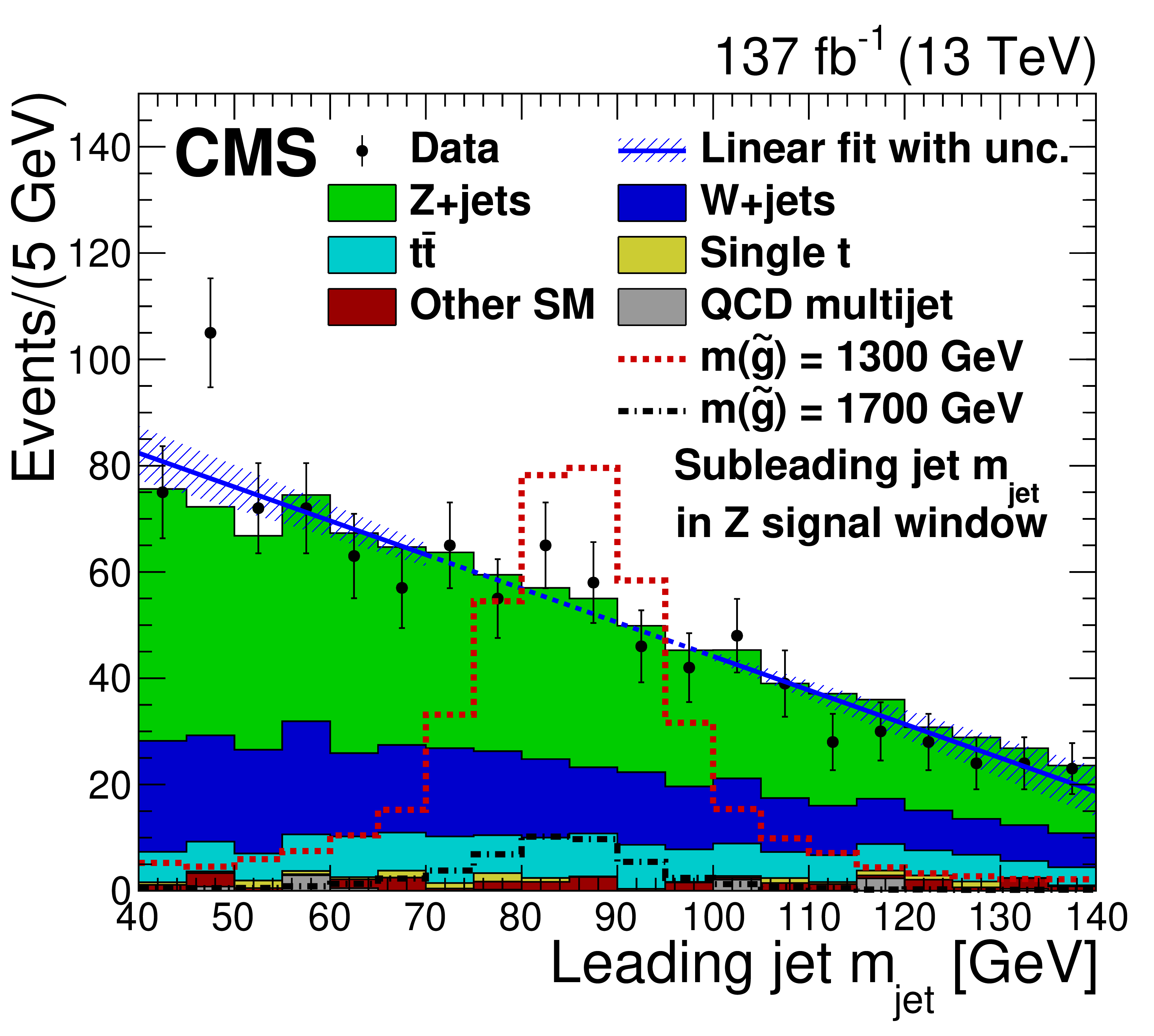} \\
\includegraphics[align=c, width=0.38\textwidth]{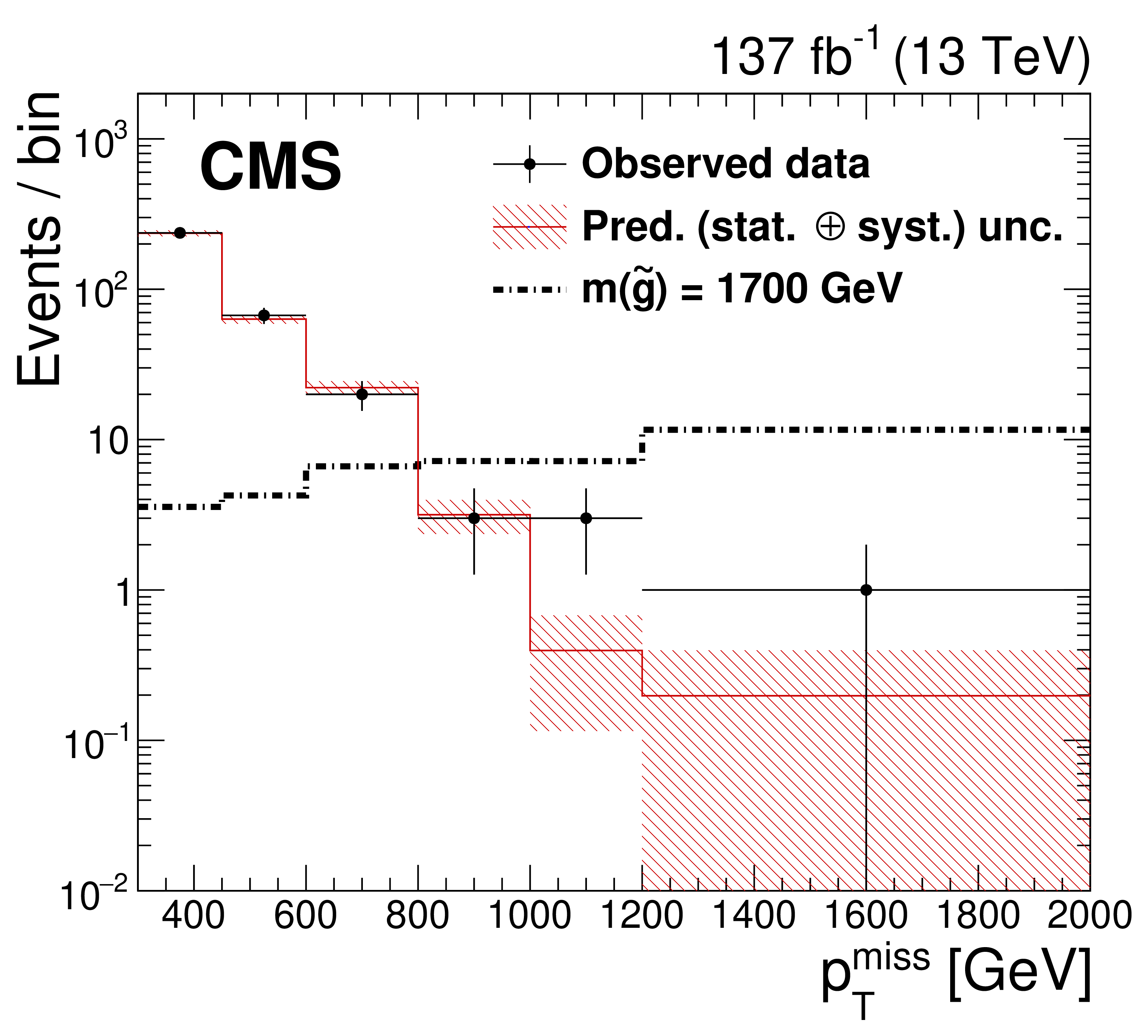} \qquad
\includegraphics[align=c, width=0.45\textwidth]{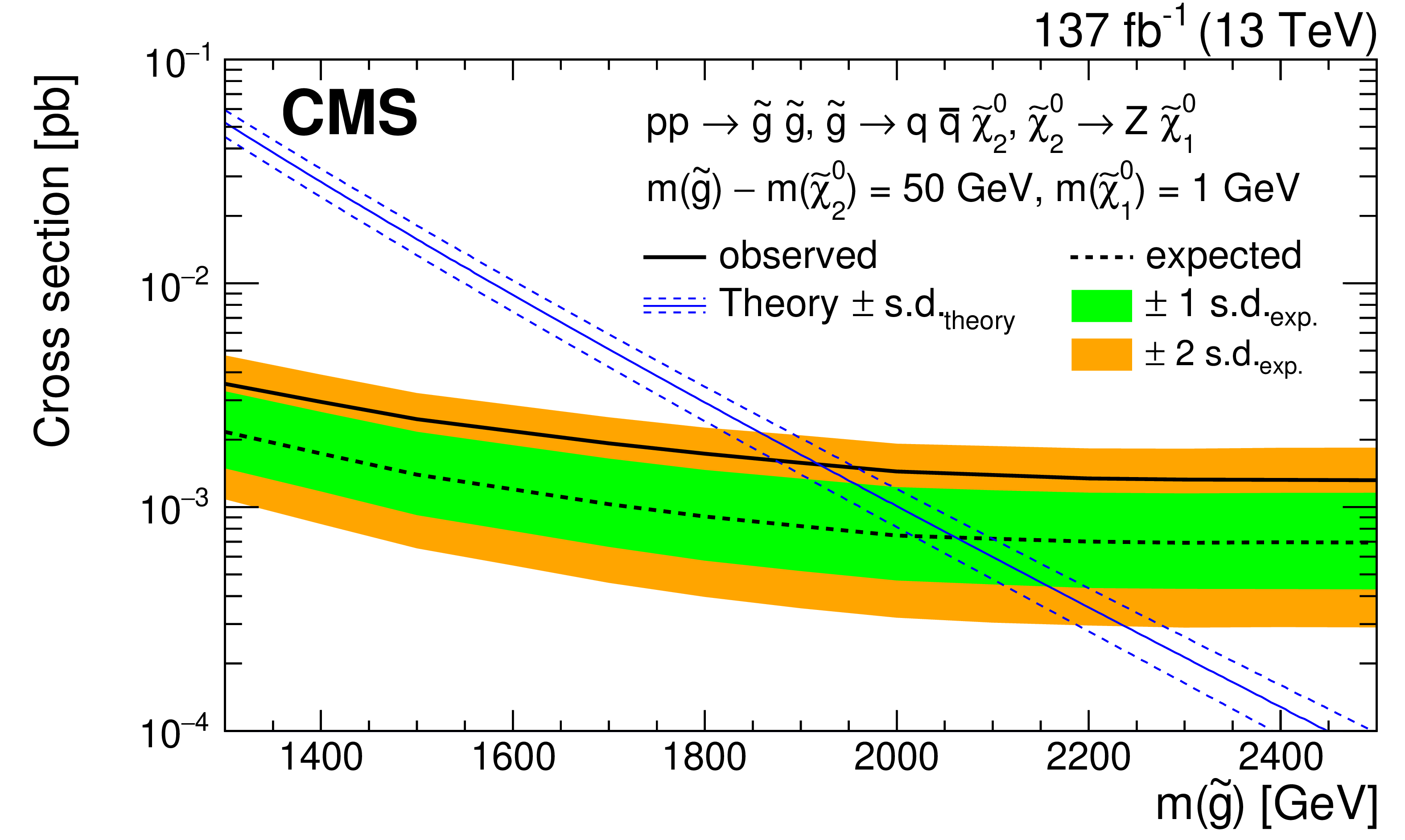}
\caption{Definition of the search and control regions in the plane of subleading vs. leading jet mass (top left), leading AK8 jet mass shape fit in the mass sidebands (top right), observed data and background prediction as functions of $p_T^{miss}$ (bottom left), and 
the 95\% CL upper limit on the production cross section for the gluino signal model as a function of the gluino mass (bottom right) in the boosted $ZZ + p_T^{miss}$ search~\cite{Sirunyan:2020zzv}.}
\label{fig:SUS-19-013}
\end{figure}

The dominant SM background in this final state is the $Z(\rightarrow \nu \nu)+$jets process.  SM backgrounds are estimated directly from data, in control regions defined using the masses of the $Z$ candidate jets as seen in Figure~\ref{fig:SUS-19-013}, top left.  The mass sideband control regions are used to fit the leading AK8 jet mass distribution for estimating the background normalization integrated over $p_T^{miss}$, as seen in Figure~\ref{fig:SUS-19-013}, top right.  The $p_T^{miss}$ control region is used to derive the $p_T^{miss}$ shape, based on the assumption that jet mass and $p_T^{miss}$ have minimal correlation.  

The estimated background yields are shown as a function of $p_T^{miss}$ and compared to data, as shown in Figure~\ref{fig:SUS-19-013}, bottom left.  No excess over the SM expectation is observed.  The results are interpreted using a simplified SUSY model of gluino pair production in which the gluino decays to a low momentum quark pair and $\nt_2$, and $\nt_2$ decays to a boosted $Z + \nt_1$, where $m_{\tilde{g}} - m_{\nt_2} = 50$~GeV and $m_{\nt_1} = 1$~GeV, as shown in Figure~\ref{fig:SUS-19-013}, bottom right.  For this scenario, data exclude gluino masses below 1920 GeV at 95\% confidence level.

%%%%%%%%%%

%\section{Compressed SUSY with soft taus}

{\bf Compressed SUSY search with soft taus: }
The second search targets directly or indirectly produced staus with low $m_{\tilde{\tau}} - m_{\nt_1}$ ($< 50$~GeV), a compressed case favored by dark matter coannihilation scenarios, where the coannihilation between the stau and the lightest neutralino can generate the observed relic density.  It is the first LHC search for a signature of one soft, hadronically decaying $\tau$ lepton ($\tau_h$), one energetic jet from initial-state radiation (ISR), and large transverse momentum imbalance~\cite{Sirunyan:2019mlu}.  The search, using 77 \fbi~of 13~TeV data, selects events having exactly one $\tau_h$ with $20 < p_T(\tau_h) < 40$~GeV, an ISR jet with $p_T > 100$~GeV, $p_T^{miss} > 230$~GeV, angular separation between the ISR jet and $p_T^{miss}$ $\Delta R(j_{ISR}, p_T^{miss}) > 0.7$ and zero $b$-jets.  The analysis looks for an excess in the distribution of $\tau$ transverse mass~$m_T(\tau_h, p_T^{miss}) = \sqrt{2p_T^{miss}p_T(\tau_h)(1 - \cos\Delta\phi( \vec{p}_T^{miss}, \tau_h) ) }$.
%\begin{equation}
%m_T(\tau_h, p_T^{miss}) = \sqrt{2p_T^{miss}p_T(\tau_h)(1 - \cos\Delta\phi( \vec{p}_T^{miss}, \tau_h) ) }
%\end{equation}

\begin{figure}[htbp]
\begin{center}
\includegraphics[width=0.48\textwidth]{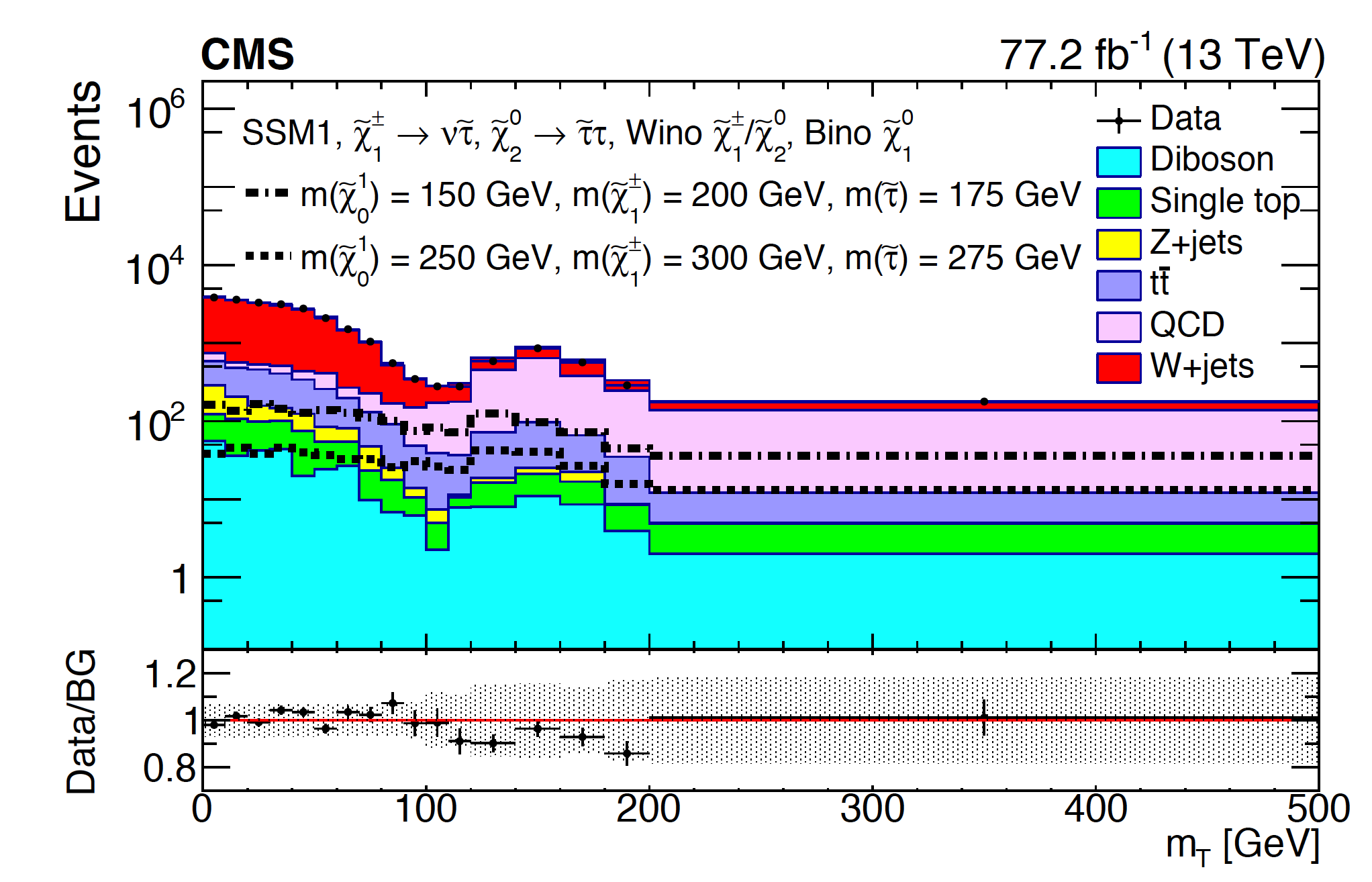} \\
\includegraphics[width=0.48\textwidth]{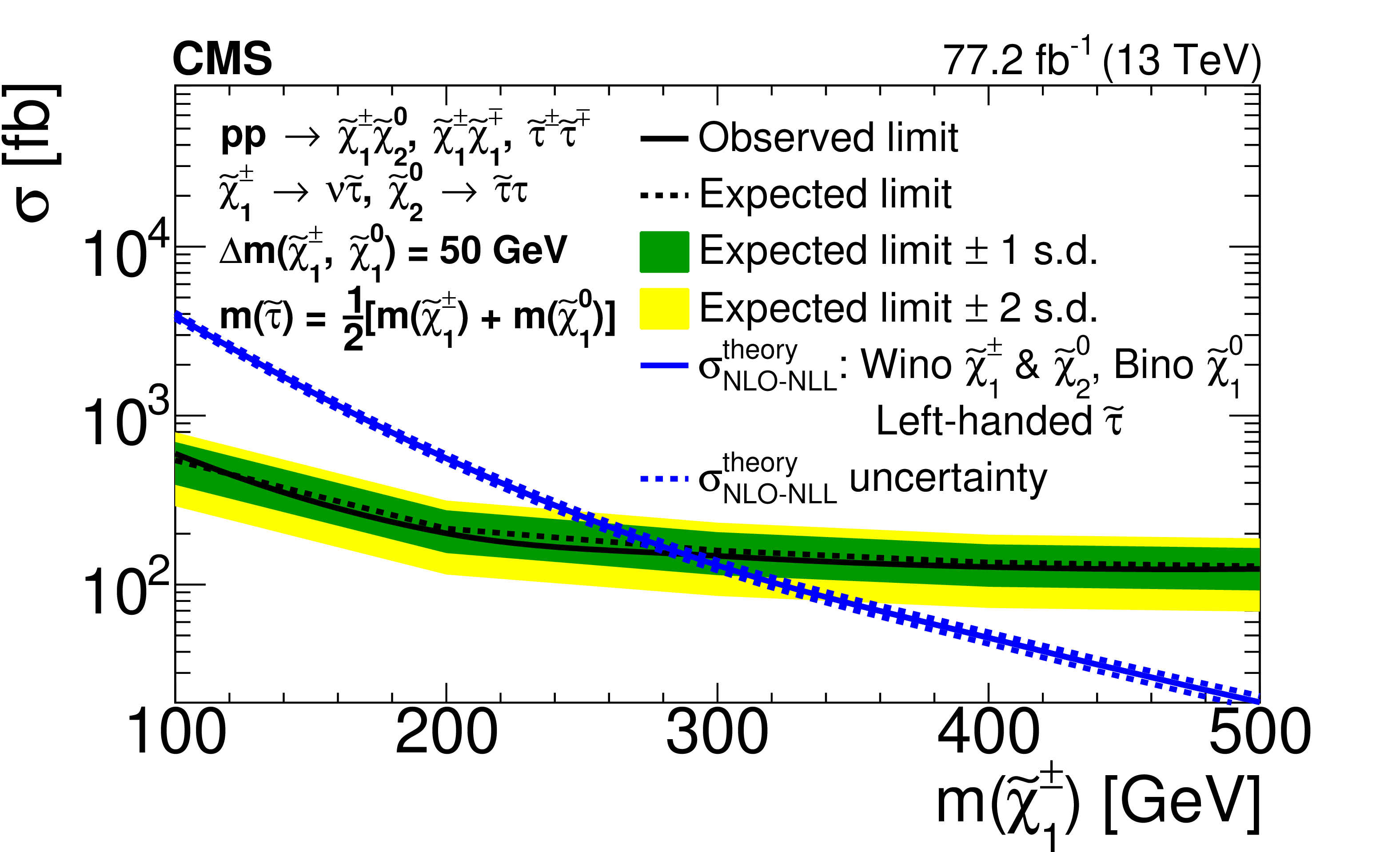}
\includegraphics[width=0.48\textwidth]{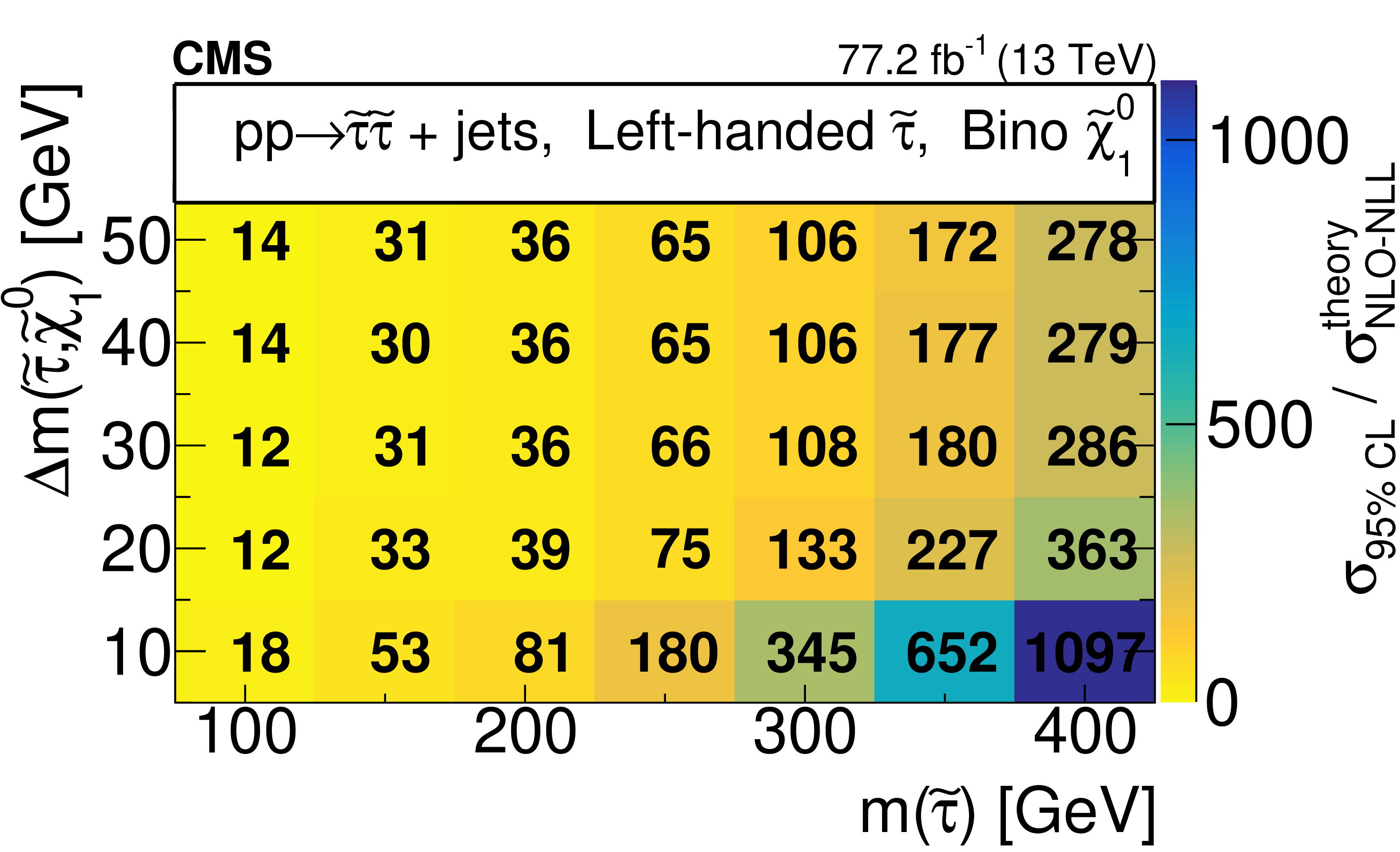}
\caption{The $m_T$ distribution of data, background prediction and signal benchmarks in the signal region (top); the 95\% CL upper limits on the $\nt_2 \ch_1 / \tilde{\chi}_1^+ \tilde{\chi}_1^-$ production cross sections as a function of $m(\tilde{\chi}_1^\pm)$ (bottom left); and  ratio of the 95\% CL upper limit on the direct $\tilde{\tau}$ pair production cross section to the theory prediction as function of $m(\tilde{\tau})$ and $\Delta m(\tilde{\tau}, \tilde{\chi}_0^1)$ (bottom right) in the compressed SUSY search with soft taus~\cite{Sirunyan:2019mlu}.}
\label{fig:SUS-19-002}
\end{center}
\end{figure}

The dominant SM backgrounds are $tt+$jets and $W/Z+$jets.  Their $m_T$ shapes are estimated from control regions and yields are extrapolated from simulation. 
 Data-simulation agreement in control regions is used to validate the modeling of the $\tau_h$ selections and to measure data-tosimulation scale factors to correct ISR jet and the $p_T^{miss}$ modeling.   
For QCD multijet backgrounds, both $m_T$ shape and yields are estimated from data control regions.  The resulting $m_T$ distribution is shown in Figure~\ref{fig:SUS-19-002}, top, where data are seen to be consistent with the SM.  Figure~\ref{fig:SUS-19-002} (bottom left) shows the interpretation of this result in a simplified SUSY model of $\nt_2 \ch_1 / \tilde{\chi}_1^+ \tilde{\chi}_1^-$ production.   For 100\% wino $\nt_2 / \chi_1$, $m_{\ch_1} - m_{\nt_1} = 50$~GeV, $m_{\tilde{\tau}} = \frac{1}{2}(m_{\ch_1} + m_{\nt_1})$ and BR($\ch_1 \rightarrow \tilde{\tau} \nu_\tau \rightarrow \tau \nt_1 \nu_\tau) = 100\%$, $\nt_2 / \ch_1$ masses up to 290 GeV are excluded at 95\% confidence level. This sensitivity exceeds that of all other searches to date, including the LEP exclusion of $m_{\chi_1} > 103.5$~GeV in compressed scenarios~\cite{lepchargino}.  Figure~\ref{fig:SUS-19-002}, bottom right shows the ratio of the 95\% CL upper limit on the direct $\tilde{\tau}$ production signal cross section to the theoretical cross section as a function of $m_{\tilde{\tau}}$ and $\Delta  m (\tilde{\tau}, \nt_1)$.  No sensitivity is achieved to direct stau pair production yet.

%\section{Disappearing track search using $M_{T2}$}

{\bf Disappearing track search using $M_{T2}$: }
For compressed SUSY with $m_{\ch_1} - m_{\nt_1} \sim O(100~MeV)$, $\ch_1$ is long lived.  It would decay in the CMS tracker to a soft, undetectable pion and a $\nt_1$.  This would lead to a disappearing track $+ E_T^{miss}$ signature.  The final search presented here looks in 137 \fbi~of 13~TeV data for such compressed charginos in gluino or squark decays by extending the classical inclusive hadronic search based on the stransverse mass variable $M_{T2}$ with final states consisting of disappearing tracks (DTs) and at least 2 jets and $M_{T2} > 200$~GeV~\cite{Sirunyan:2019xwh}.  The search explores categories of short and medium/long DT selections, which consist of hits in the pixel or pixel $+$ strip tracking detectors of CMS, respectively, in order to search for a wide range of lifetimes.  Including DTs in the search gives a possibility to loosen kinematic requirements without accumulating large amounts of backgrounds.  For instance, the  $M_{T2}$ requirement is reduced from 400 to 200~GeV.  The analysis categorizes events in 68 search bins defined in jet multiplicity, hadronic transverse momentum $H_T$, DT length and DT $p_T$.  Main sources of backgrounds are hadrons and leptons poorly reconstructed in the tracker and tracks built out of incorrect combinations of hits.  They are estimated by calculating fake rates in data control regions and applying these fake rates to DT candidates.

\begin{figure}[htbp]
\begin{center}
\includegraphics[width=0.32\textwidth]{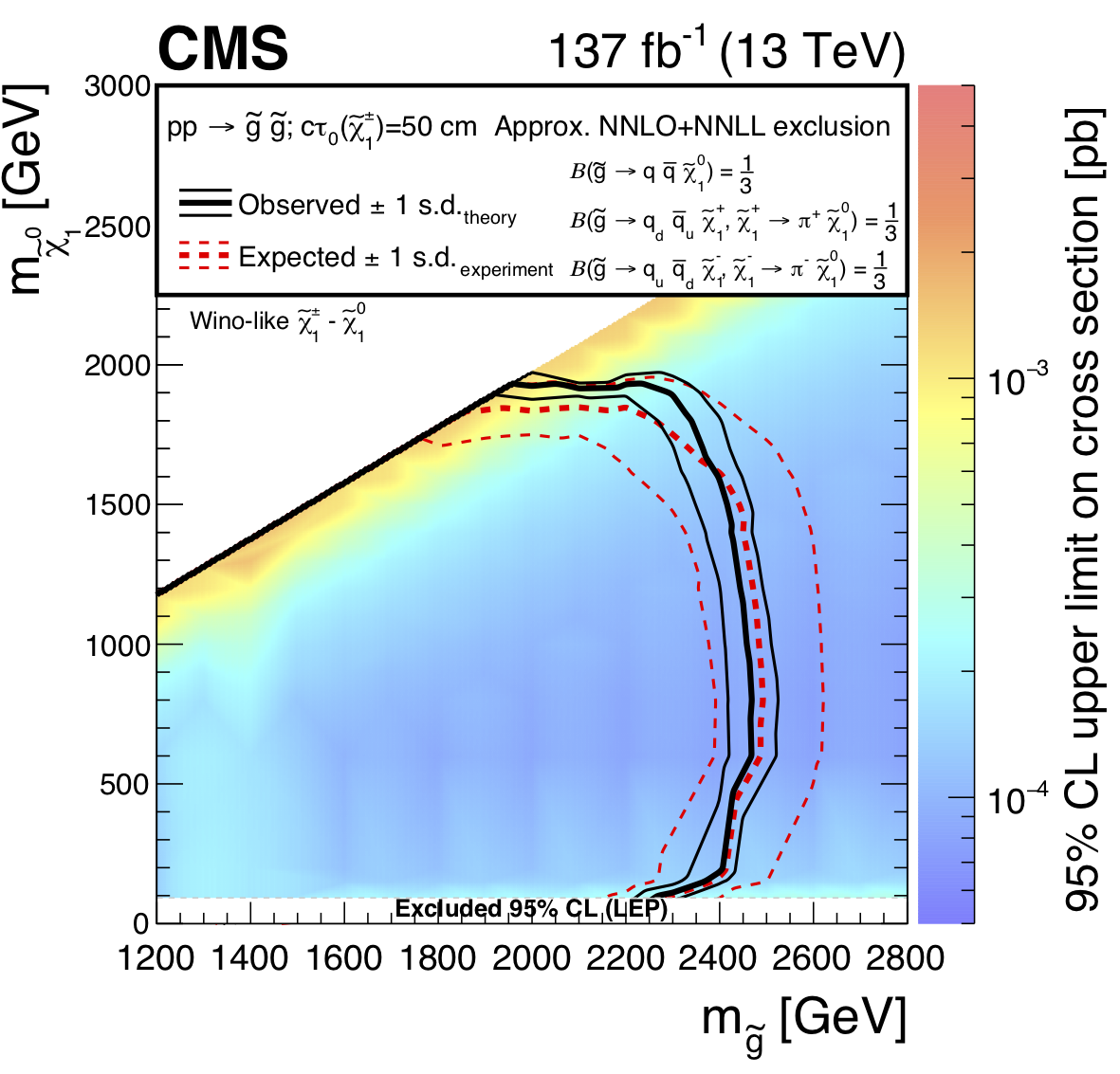}
\includegraphics[width=0.32\textwidth]{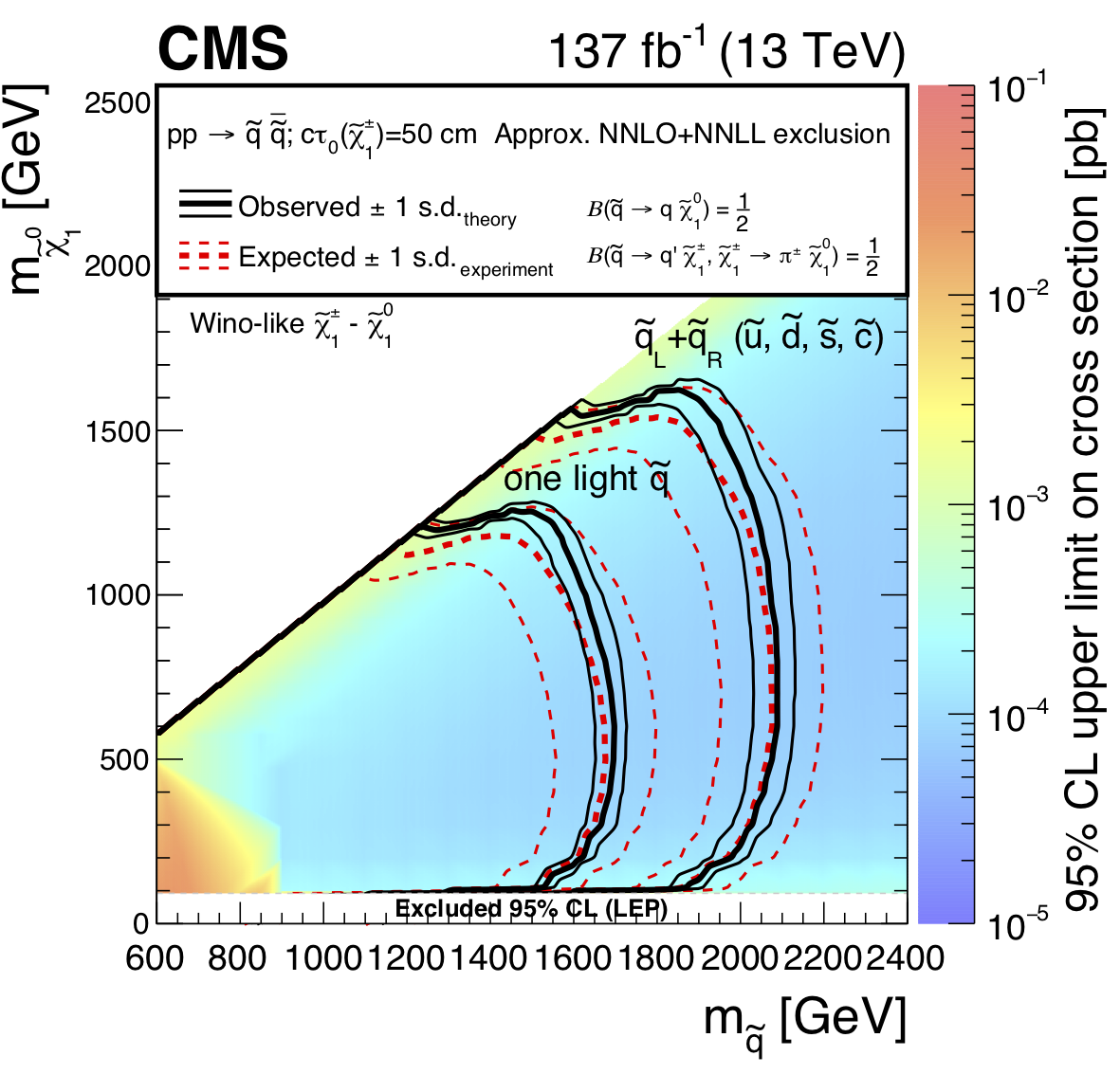}
\includegraphics[width=0.32\textwidth]{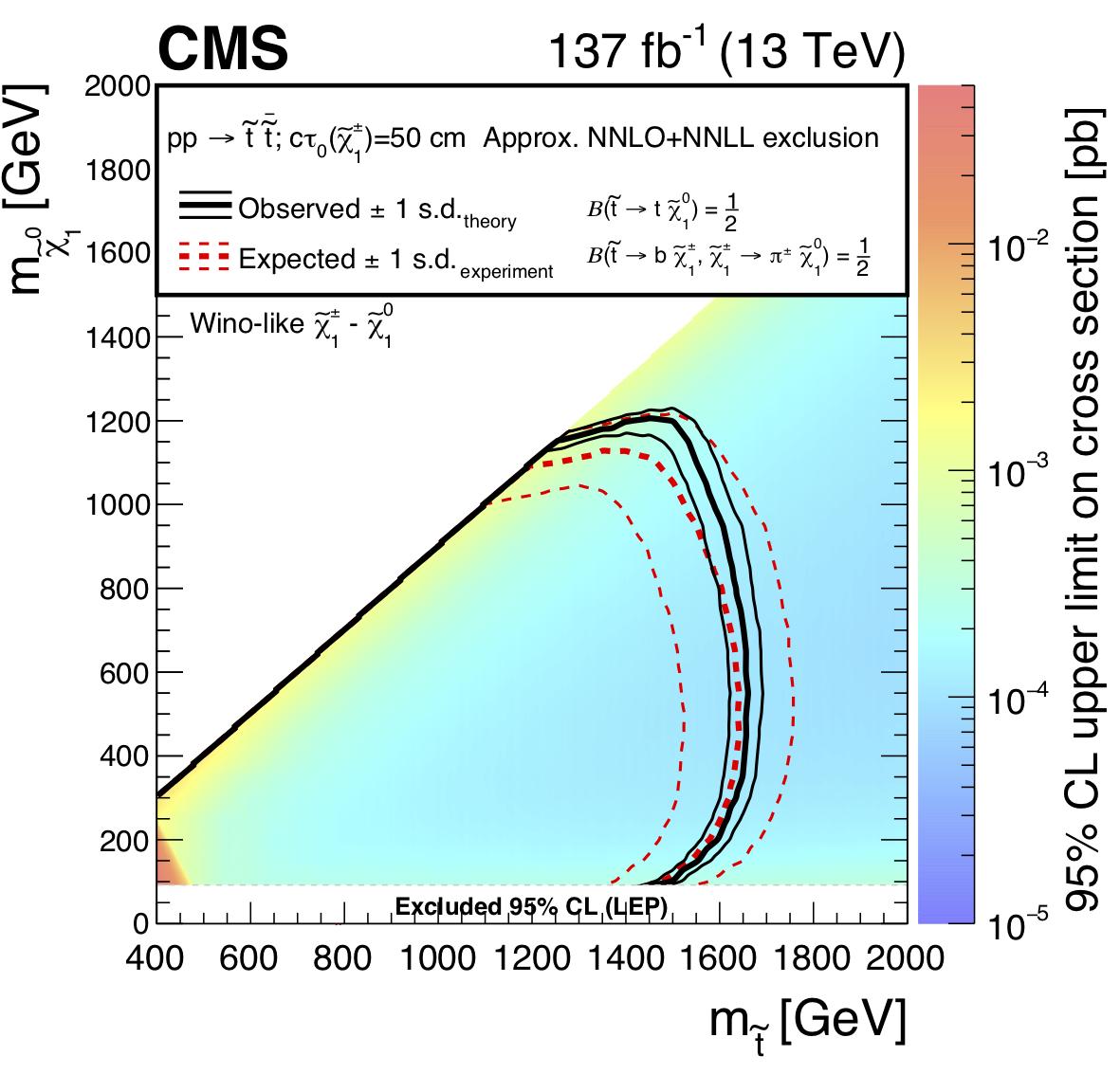}
\includegraphics[width=0.32\textwidth]{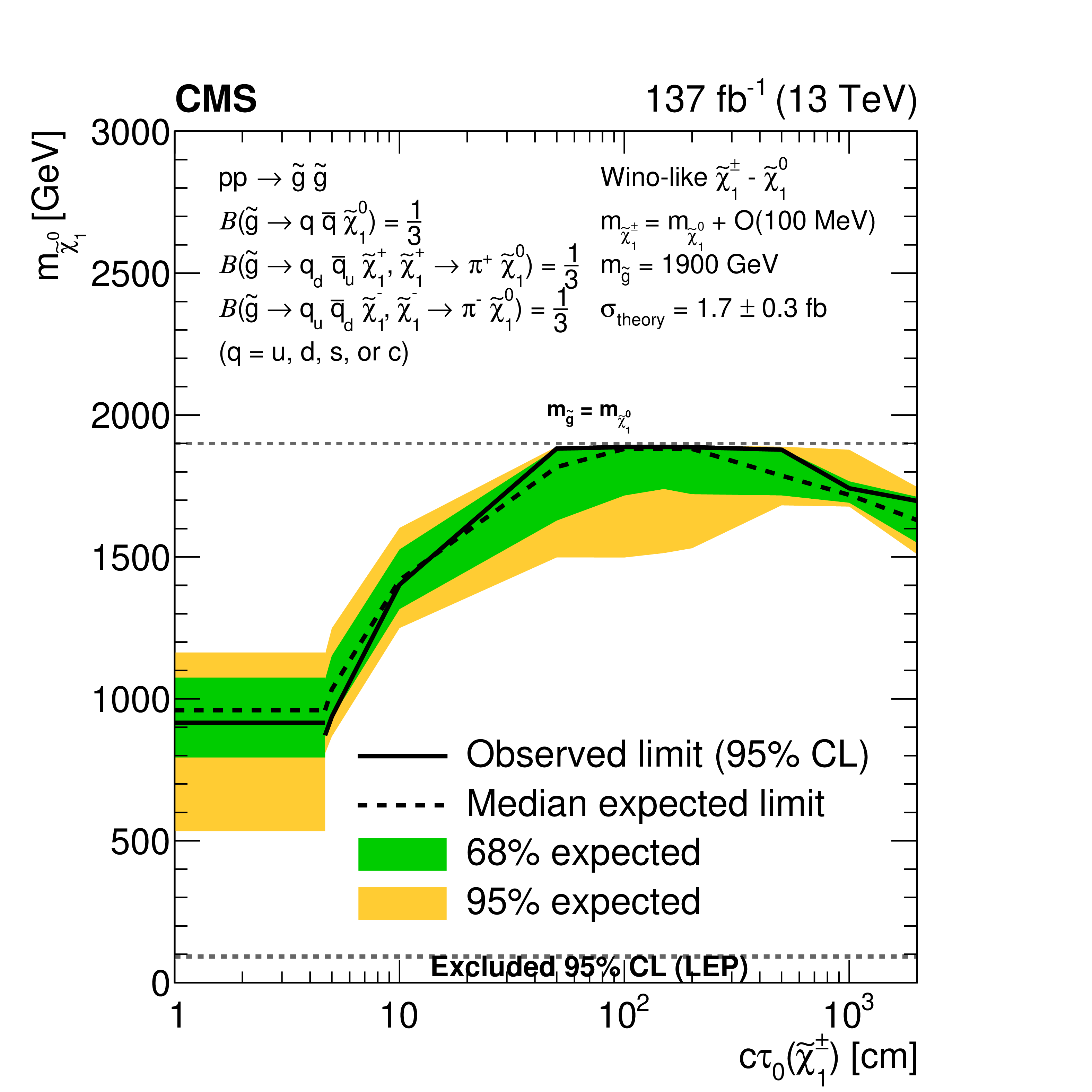}
\includegraphics[width=0.32\textwidth]{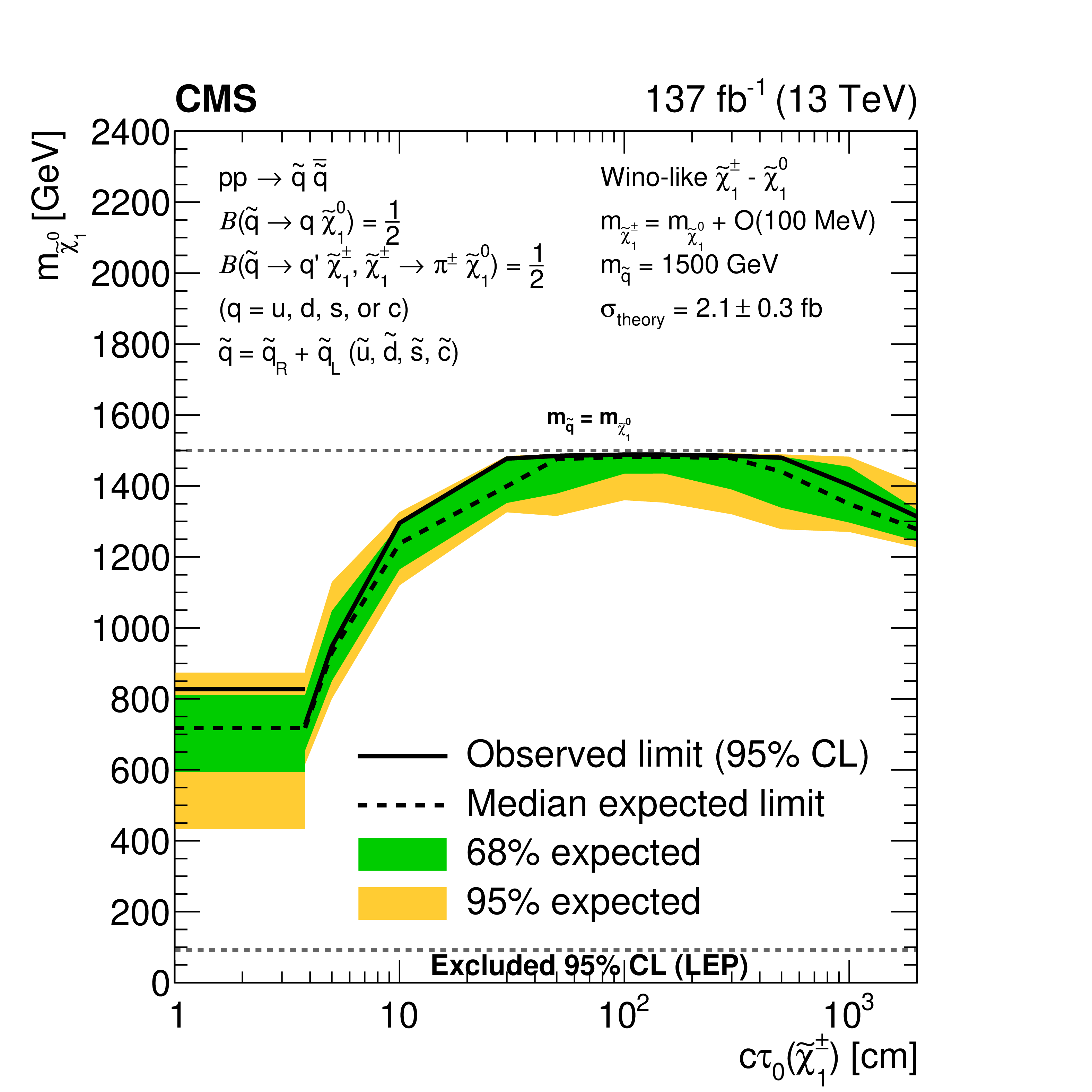}
\includegraphics[width=0.32\textwidth]{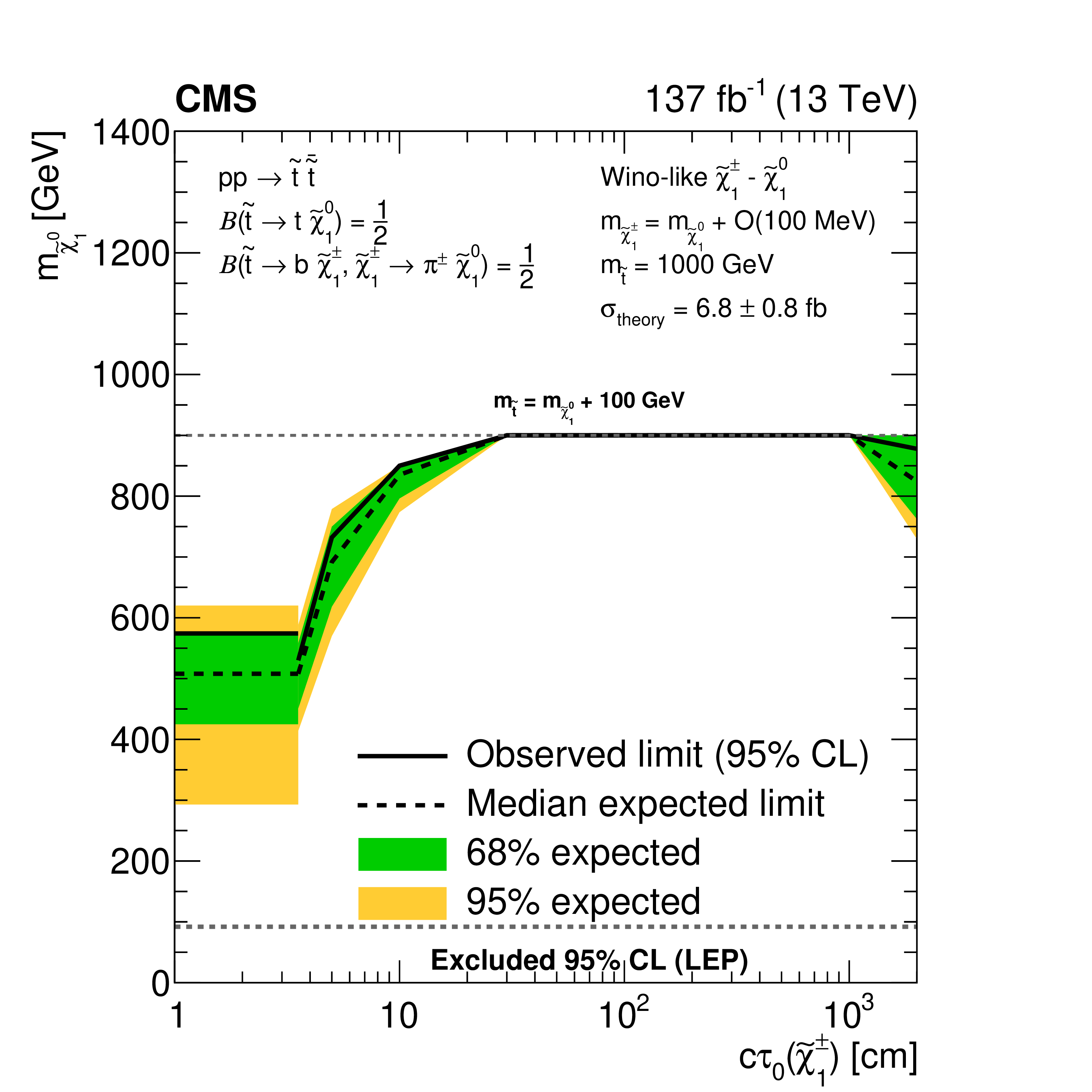}
\caption{Exclusion limits at 95\% CL for direct gluino pair production where the gluinos decay to light-flavor quarks (top left), light squark pair production (top center), and top squark pair production (top right) with $c\tau_0(\tilde{\chi}^\pm_1) = 50$~cm.  Exclusion limits on $m_{\nt_1}$ with $m_{\ch_1} = m_{\nt_1} + O(100 MeV)$ as a function of $\ch_1$ proper decay length for gluino pair production with $m_{\tilde{g}} = 1900$~GeV (bottom left), squark pair production with $m_{\tilde{q}} = 1500$~GeV (bottom center), and top squark pair production with $m_{\tilde{t}} = 1000$~GeV (bottom right) in the disappearing track search using $M_{T2}$~\cite{Sirunyan:2019xwh}.}
\label{fig:SUS-19-005}
\end{center}
\end{figure}

The search found no deviation in data from the SM expectation.  Figure~\ref{fig:SUS-19-005} shows the interpretation of the search results in various simplified SUSY models.  The top row shows exclusion limits at 95\% CL for direct gluino pair production where the gluinos decay to light-flavor (u, d, s, c) quarks (top left), light squark pair production (top center), and top squark pair production (top right) for $c\tau_0(\tilde{\chi}^\pm_1) = 50$~cm.  Extending the inclusive $M_{T2}$ search with disappearing tracks increased $m_{\tilde{g}}$ reach from $\sim$2 to $2.46$~TeV and $m_{\nt_1}$ reach from $\sim$1.2 to $\sim$2~TeV for gluino pair production; $m_{\tilde{q}}$ reach from $\sim$1.8 to $\sim$2.1~TeV and $m_{\nt_1}$ reach from $\sim$0.8 to $\sim$1.6~TeV for squark pair production, and $m_{\tilde{t}}$ reach from $\sim$1.2 to 1.65~TeV and $m_{\nt_1}$ reach from $\sim$0.55 to $\sim$1.25~TeV for stop pair production.  In all cases, sensitivity in the compressed region was significantly improved.  The bottom row in Figure~\ref{fig:SUS-19-005} shows exclusion limits versus chargino decay length for selected gluino, squark and stop masses.% for gluino, light squark and top squarks with masses 1900, 1500 and 1000~GeV respectively.

%\section{Summary and outlook}

In summary, 3 recent examples of dedicated CMS SUSY searches targeting specific scenarios and exclusive signatures were presented, namely, a boosted $ZZ + p_T^{miss}$ search, which extended the gluino mass reach to 1.9 TeV for $m_{\tilde{g}} - m_{\nt_2} = 50$~GeV; a soft hadronic $\tau + p_T^{miss} + $ISR jet search for compressed staus motivated by dark matter coannihilation models, which obtained a sensitivity for charginos extending the LEP limits; and a search that added regions with disappearing tracks to the inclusive hadronic $M_{T2}$ search, which increased gluino and squark mass limits by 400-600 GeV and significantly improved sensitivity in the compressed region.  Other searches dedicated to specific final states have been performed earlier, such as the search for $H \rightarrow gg$ using razor and $M_{T2}$ variables~\cite{CMS:2019eln}; searches for hadronic and semileptonic staus~\cite{CMS-PAS-SUS-17-002, Sirunyan:2018vig}; $E_T^{miss}$ and boosted Higgs to $bb$~\cite{Sirunyan:2017bsh}; SUSY in vector boson fusion channels~\cite{Sirunyan:2019zfq}; RPV smuons~\cite{Sirunyan:2018hwm}; selectrons and smuons~\cite{Sirunyan:2018nwe}; the searches in diphoton and $E_T^{miss}$ final states~\cite{Sirunyan:2019mbp}, $b$ jets and photons final states~\cite{Sirunyan:2019hzr}, and photon, lepton and $E_T^{miss}$ final states~\cite{Sirunyan:2018psa}. More searches are ongoing for soft opposite-sign 2 lepton signatures and steath and RPV stops.  
%Other searches dedicated to specific final states have been performed earlier, such as the search for $H \rightarrow gg$ using razor and $M_{T2}$ variables; searches for hadronic and semileptonic staus;  $E_T^{miss}$ and boosted Higgs to $bb$, SUSY in vector boson fusion channels; RPV smuons; selectrons and smuons, the searches in diphoton and $E_T^{miss}$ final states $b$ jets and photons final states and photon, lepton and $E_T^{miss}$ final states~\cite{others}. More searches are ongoing for soft opposite-sign 2 lepton signatures and steath and RPV stops.  

\vspace{0.3cm}
{\bf Acknowledgements: } I would like to thank my colleagues in the CMS Collaboration for their hard work in producing the results in this contribution, and to the organizers of ICHEP 2020 for their efforts in realizing this important conference virtually during the difficult Covid-19 period.

\end{document}